\documentclass[a4paper]{spie}  %>>> use this instead for A4 paper

\usepackage[left=1.925cm,top=2.54cm,right=1.925cm,bottom=4.94cm,bindingoffset=0.0cm]{geometry}
\usepackage[hyphens]{url}
\usepackage{amsmath,amsfonts,amssymb}
\usepackage{graphicx}
\usepackage{braket}
\usepackage{hyperref}
\hypersetup{colorlinks=true, allcolors=blue}

\newcommand*\circled[1]{%
        \raisebox{.5pt}{\textcircled{\raisebox{-0.4pt} {{\sffamily {\scriptsize #1}}}}}%
}

\title{The European Satellite-Based QKD System EAGLE-1}

\author{Thomas~Hiemstra}
\author{David~Hasler}
\author{Domenico~Paone}
\author{Fabian~Reichert}
\author{Frank~Heine}
\author{Julian~Struck\thanks{~~julian.struck@tesat.de}}
\affil{Tesat-Spacecom GmbH \& Co. KG, Laser Systems Development,\\Gerberstra{\ss}e 49, Backnang, Germany}

\begin{document} 
\maketitle

\begin{abstract}
The satellite mission EAGLE-1 represents an important step towards a future pan-European secure quantum key distribution (QKD) network. The public-private partnership behind the mission consists of a consortium of universities, research institutes, and companies partially funded by ESA, the European Union, and supported by national delegations. This unique combination of academic partners and industry facilitates a swift knowledge transfer from basic research to commercial application. Within the consortium, Tesat-Spacecom (TESAT) is responsible for developing and integrating the payload assembly of the low-earth orbit EAGLE-1 satellite. In addition, TESAT provides the SCOT80 laser terminal with minor adaptations for the mission. Here, we report on the status, technical aspects and TESATs contribution to the satellite-to-ground prepare-and-measure QKD mission.
\end{abstract}


\section{INTRODUCTION}
\label{sec:intro}
Quantum key distribution (QKD) describes a whole class of cryptographic symmetric-key exchange protocols that utilize fundamental principles of quantum physics\cite{xu_secure_2020,pirandola_advances_2020,wolf_quantum_2021}. All of these protocols have in common that  the cryptography key material is encoded as qubits onto light pulses in the single- or few-photon regime. An eavesdropper intercepting and measuring these qubits leaves a detectable statistical trace on exchanged keys. The same argument applies if an eavesdropper tries to duplicate or amplify the qubits. Therefore, if a shared key displays no statistically significant anomaly, its privacy and integrity is guaranteed.

While the laws of quantum mechanics protect the key exchange, they also create novel technical challenges not encountered in classical communication systems. A major issue for the practical use of QKD are quantum channel losses. Since amplification of the quantum signal is not an option and the current technological maturity of quantum repeaters is too low\cite{munro_inside_2015,azuma_quantum_2023}, photon losses cannot be compensated. As a result, the shared key material per time, the so-called key rate, decreases with increasing channel losses until it reaches the system specific noise floor and key sharing becomes impossible. For terrestrial QKD via optical fibers, where the signal attenuation scales exponentially with the fiber length, the maximum link distance is currently limited to a few hundred kilometers \cite{pittaluga_600_2021,chen_twin-field_2021,clivati_coherent_2022,chen_quantum_2022,zhou_twin-field_2023}. This is sufficient to establish secure key exchange in local networks but a global communication architecture mandates key generation over significantly longer distances.

Here, spaceborne free-space optical (FSO) links can complement the terrestrial fiber infrastructure to enable long-range QKD. Optical satellite-to-ground links feature significantly lower propagation losses than fiber links. Under good weather conditions, the main contributor to losses is typically laser beam divergence. In the far-field, these diffraction losses scale quadratically with distance which facilitates significantly longer QKD link distances when compared to fiber links. The feasibility of satellite-to-ground QKD has already been successfully demonstrated with the quantum satellite Micius, launched in 2016 as part of the Chinese Quantum Experiments at Space Scale research project \cite{lu_micius_2022}. Further in-orbit QKD missions have demonstrated QKD payloads with reduced size, weight and power consumption (SWaP)\cite{liao_space--ground_2017,li_spaceground_2022,li_microsatellite-based_2024} and integration with terrestrial QKD fiber networks \cite{chen_integrated_2021}. This paves the way towards QKD satellite constellations and global QKD coverage.

\section{Mission Overview \& Key Facts}

EAGLE-1 is a public-private partnership project with the objective to develop, deploy and operate Europe's first spaceborne prepare-and-measure QKD system and to serve as a pathfinder for a future European QKD constellation \cite{noauthor_eagle-1_nodate,noauthor_eagle-1_2024}. It consists of an SES-led consortium with 20 partners from academia, public research institutes and industry and is co-funded by ESA, the European Union (EU) and the space agencies of Germany, Luxembourg, Austria, Italy, the Netherlands, Switzerland, Belgium and the Czech Republic, as well as the industry. The mission encompasses a space segment with a satellite in a sun-synchronous low earth orbit, a ground segment with two optical ground stations (OGSs), and a coordinating operations control center. The satellite is currently scheduled for launch in 2026 by Arianespace on-board a Vega C rocket from the Guiana Space Centre.

\begin{figure} [b]
\begin{center}
\includegraphics[width=\linewidth]{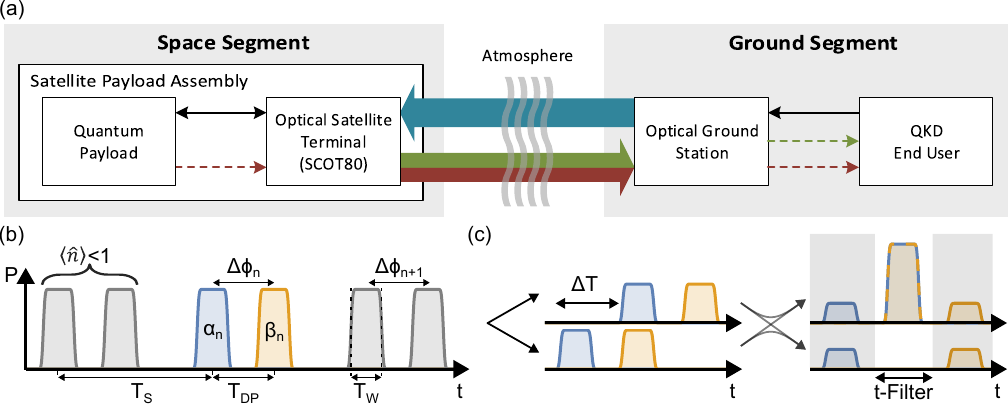}
\end{center}
\caption[EAGLE-1 communication architecture.] 
{ \label{fig:E2E_Phase} 
QKD architecture and quantum signal. (a) High-level overview of the QKD architecture. Black solid arrows represent electronic interfaces, dashed arrows correspond to single-mode fiber links, while the thick arrows indicate FSO links. Red, green, and blue colors label the quantum channel ($\lambda_\mathrm{Q}\approx 1565.5\,\mathrm{nm}$), public down-link channel ($\lambda_\mathrm{DL}\approx 1553.3\,\mathrm{nm}$), and public up-link channel ($\lambda_\mathrm{UL}\approx 1536.6\,\mathrm{nm}$), respectively. (b) Double pulse phase encoding scheme for quantum states with $\Delta \phi_{n}=\beta_{n}-\alpha_{n}$ [Fig. \ref{fig:E2E_Phase}(b)]\cite{gunthner_eagle-1_2024}. The double pulse repetition period is $T_{\mathrm{S}}=400\,\mathrm{ps}$ with a single pulse width of $T_{\mathrm{W}}=80\,\mathrm{ps}$ and a temporal spacing of $T_{\mathrm{DP}}=160\,\mathrm{ps}$. Not shown here are weak decoy states and bright reference pulses. (c) Delayed self-homodyne phase decoding scheme.
Double pulses are split, relatively time shifted by $\Delta T$ and then recombined causing interference that is converted into counts on single-photon detectors that are time filtered in post processing \cite{hacker_phase-locking_2023,gunthner_eagle-1_2024}.}
\end{figure}

A key design driver for the EAGLE-1 system is compatibility with existing telecommunication hardware and infrastructure in order to simplify a future integration into existing fiber networks and reduce adaptation costs. As a consequence, all required communication channels for the secure key exchange are located in the optical C-band, i.e. the quantum channel and the  public (classical) down-link and up-link channel [Fig. \ref{fig:E2E_Phase}(a)]. To facilitate real-time key distillation and QKD post processing, all communication channels are used simultaneously. The exchanged secret key is stored at the QKD end user and also temporarily in the quantum payload (QPL). In this respect, the satellite or more precisely the QPL has to be treated as a trusted node in a QKD network and needs to be protected accordingly.

The EAGLE-1 specific QKD protocol, developed by researchers under the lead of the Max Planck Institute for the Science of Light (MPL) and the University of Erlangen-Nuremberg (FAU), is a discrete-variable decoy-state BB84 variant \cite{bennett_quantum_1984,hwang_quantum_2003,gunthner_eagle-1_2024}. This type of protocol tolerates losses sufficiently well for LEO-satellite QKD, has a high level of maturity with respect to security proofs and the required technology is readily available \cite{orsucci_assessment_2024}. For the information transmission on the quantum channel, 4 qubit states are used that form two mutually unbiased orthonormal bases $\mathcal{B}_{z}=\lbrace\ket{0},\ket{1}\rbrace$ and $\mathcal{B}_{x}=\lbrace\ket{0^\prime},\ket{1^\prime}\rbrace$ with $\ket{0^\prime}=\left(\ket{0}+\ket{1}\right)/\sqrt{2}$ and $\ket{1^\prime}=\left(\ket{0}-\ket{1}\right)/\sqrt{2}$. These states are phase encoded onto weak coherent double pulses with a mean photon number $\braket{\hat{n}}<1$ and a temporal spacing of $T_{\mathrm{DP}}=160\,\mathrm{ps}$ via the relation $\lbrace\ket{0},\ket{1},\ket{0^\prime},\ket{1^\prime}\rbrace \leftrightarrow \Delta\phi = \lbrace 0,\pi,\pi/2,3\pi/2\rbrace$ [Fig. \ref{fig:E2E_Phase}(b)]\cite{gunthner_eagle-1_2024}. Phase-encoded QKD is directly compatible with transmission in optical fibers and has no special requirements on free-space optical systems such as telescopes. 

The inverse process of phase decoding, conducted by the QKD end user, is based on delayed self-homodyne single-photon detection [Fig. \ref{fig:E2E_Phase}(c)] \cite{vallone_interference_2016,gunthner_eagle-1_2024}. The relative delay $\Delta T = T_{\mathrm{DP}} + \phi_{\mathrm{R}}/\omega$ creates a temporal overlap of the first and second part of the double pulse and introduces a controlled relative receiver phase shift $\phi_{\mathrm{R}}$. For $\phi_{\mathrm{R}}=0\,(\pi/2)$, the measurement basis corresponds to $\mathcal{B}_{z}$ ($\mathcal{B}_{x}$), respectively. In post processing single-photon detection events are filtered to the relevant time windows to only gather interference counts.

The time synchronization between space and ground segment is based on bright references signals which are approximately $40\,\mathrm{dB}$ more intense than the quantum states \cite{rosler_eagle-1_2024}. The effective duty cycle of these reference signals is $10\,\%$ which reduces the effective time-averaged qubit symbol rate to $2.25\,\mathrm{GS/s}$. Under good atmospheric conditions, end-to-end quantum channel losses range from $40\,\mathrm{dB}$ to $60\,\mathrm{dB}$ during a satellite overpass. This translates into expected maximum secret key rates in the $\mathrm{kbit/s}$ regime for night-time operation.

\section{Space Segment}
\begin{figure} [b]
\begin{center}
\includegraphics[width=\linewidth]{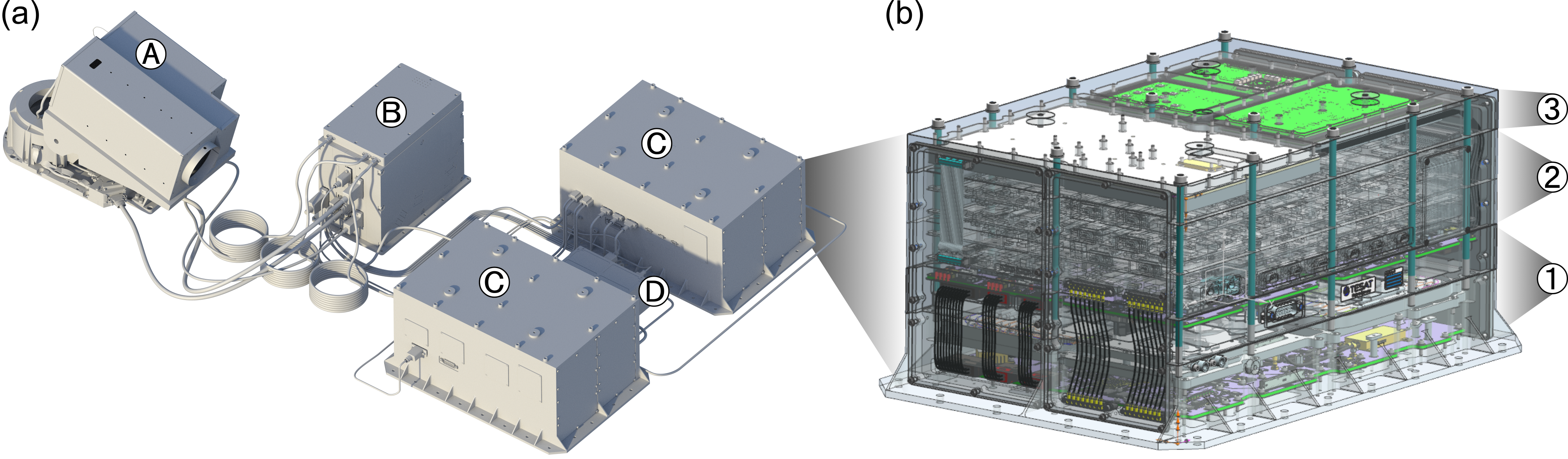}
\end{center}
\caption[Satellite payload assembly] 
{ \label{fig:PLA} 
Satellite payload assembly. (a) The satellite payload assembly consists of the SCOT80 OST and the QPL [Fig. \ref{fig:E2E_Phase}(a)]. The OST encompasses an optical head (OH) \circled{A} and an electronics unit (EU) \circled{B}. The QPL is comprised of two redundant QKD units \circled{C} and a connector box that serves as a switch for the electronic interfaces between the QKD units and the OST \circled{D}. In contrast, the optical interfaces of the QKD units are directly connected to the EU via separate single-mode fibers. The OH and EU are connected through two optical single-mode fiber interfaces that are separately used for transmitted signals (quantum signal and classical down-link) and received signals (classical up-link). (b) Detailed view of a QKD unit that consist of 5 layers of vertically stacked boards. The two bottom layers host the QKD transmitter with its electronics, electro-optical components and fiber optics  \circled{1}. The next two layers are occupied by the QKD processor and mass memory \circled{2}. The top layer includes two different types of quantum random number generators (QRNGs) \circled{3}.}
\end{figure}

In this section, we focus on the satellite payload assembly that provides the necessary functionality for a secure key exchange as part of the QKD protocol [Fig. \ref{fig:PLA}(a)]. It is developed under TESAT's lead and includes the optical satellite terminal (OST) and QPL. The payload assembly is hosted on the satellite platform PLATiNO from Sitael \cite{noauthor_sitael_2022}.

\subsection{Optical Satellite Terminal (OST)}
\label{sec:OST}

The OST is a SCOT80 from TESAT with minor adaptions to accommodate the multiplexing of the quantum channel. It is compatible with the Space Development Agency (SDA) Optical Intersatellite Link Standard (Version 2.1.2) and has a power consumption in the range from $30\,\mathrm{W}$ to $60\,\mathrm{W}$, depending on the operation mode. The SCOT80 has successfully undergone in-orbit validations on a variety of different satellite platforms \cite{noauthor_kepler_nodate,noauthor_sda_nodate,biesecker_sda_2024}.

\paragraph{Optical head (OH)}
The optical system of the OH consist of an optical bench, where the transmit and receive path are spatially (de-)multiplexed by means of a dichroic mirror, and a reflective off-axis telescope with an $80\,\mathrm{mm}$ aperture. The optical system is designed for the C-band. In addition, the optical bench includes steering mirrors for the fine pointing and point ahead correction. The entire optical system is located on an azimuth elevation drive (AED) enabling platform independent coarse pointing. The size and weight of the OH is $47 \times 31 \times 29  \,\mathrm{cm^3}$ and $7.1\,\mathrm{kg}$, respectively.

\paragraph{Electronics unit (EU)} The EU contains the optical transceiver module for the amplitude modulated classical optical signals (On-Off Keying Non-Return-to-Zero coding). Data rates up to 191.29 Mbit/s are supported, which constitutes a sufficient public channel bandwidth to facilitate real-time secret key distillation during a satellite overpass. In addition, fiber-optic beam combiners and spectral filters have been added to the EU to accommodate the multiplexing of the quantum signal and the classical down-link signal [Figs. \ref{fig:E2E_Phase}(a) and \ref{fig:PLA}(a)]. Special care has been taken to strongly suppress the out-of-band pollution from the classical channel into the quantum channel, which is essential given that the quantum signal is roughly 9 orders of magnitude weaker than the classical signal\footnote{The average power of the quantum signal is on the order of a few-hundred $\mathrm{pW}$ while the classical signal is in the few-hundred $\mathrm{mW}$ regime.}. The size and weight of the EU is $26 \times 16 \times 21\,\mathrm{cm^3}$ and $7.3\,\mathrm{kg}$, respectively.

\subsection{QKD Unit}
The QKD unit has been specifically developed for the EAGLE-1 mission in a co-engineering effort by several consortium members under the lead of TESAT. It includes two redundantly operating quantum random number generators (QRNGs) with different types of entropy sources, a QKD processor and a QKD transmitter [Fig. \ref{fig:PLA}(b)]. It has a power consumption in the range from $60\,\mathrm{W}$ to $190\,\mathrm{W}$, depending on the operation mode and a size and weight of $41 \times 29 \times 19\,\mathrm{cm^3}$ and $18\,\mathrm{kg}$, respectively.

\paragraph{Quantum random number generators (QRNGs)} The QKD unit features two independent QRNGs, developed by ID Quantique and OHB that utilize different quantum mechanical effects as entropy sources in order to generate random bits at rates of a few tens of $\mathrm{Mbit/s}$. The ID Quantique QRNG is based on photon number quantum fluctuations from a light-emitting diode \cite{gras_quantum_2021,simondi_id_2023}. In contrast, the OHB QRNG is based on amplified quantum fluctuations in a coherent detection scheme \cite{kobel_random_2023}.

\paragraph{QKD processor}
The QKD processor is developed by TSD Space with software contributions from the Austrian Institute of Technology, esc Aerospace and SES. The unit fulfills four main functions: payload management, QKD protocol handling, randomness buffering on a terabyte-sized mass memory and temporary secret key storage for the key relay between two QKD end users.

\paragraph{QKD transmitter}
The QKD transmitter generates the phase-encoded optical signal for the quantum channel, including signal and decoy quantum states as well as bright reference pulses. The transmitter also features a variety of monitoring systems and a quantum hacking protection \cite{rivera_building_2024}. It is developed by the DLR Institute of Communications and Navigation (DLR-IKN) and translated into a space-qualified flight model by TESAT.

\section{Ground Segment}
Conceptually, the ground segment can be separated into OGS and QKD end user [Fig. \ref{fig:E2E_Phase}(a)]. The OGS collects the laser light from the satellite, separates the quantum from the classical down-link signal and couples both into separate single-mode fibers that lead to the QKD end user\footnote{Sharing the same fiber for quantum and classical channel over extended distances leads to significant Raman scattering into the quantum channel.}. For the classical up-link signal, the QKD end user sends electronic signals to the OGS, where those electronic signals are converted into optical signals and transmitted back to the satellite.

\paragraph{Optical ground station (OGS)}
EAGLE-1 will demonstrate secret key exchange between sites in Germany and the Netherlands. The Dutch OGS is developed under the lead of TNO and Airbus\cite{}. The satellite link to Germany will be established through the recently upgraded OGS of DLR-IKN in Oberpfaffenhofen, featuring an Nasmyth telescope with an $80\mathrm{cm}$ aperture and an advanced adaptive optics system for fiber coupling \cite{rivera_building_2024}. The classical satellite down-link signal serves as a beacon for the pointing, acquisition, and tracking (PAT) system of the OGS. For the optical up-link, spatial transmitter diversity mitigates detrimental interference effects caused by atmospheric scintillation. 

\paragraph{QKD end user}
The QKD end user represents the ground segment counterpart of the QPL and secret-key recipient. The user requires a quantum receiver, classical optical modem, QKD processor and key management system. The QKD receiver, that decodes the optical quantum signal, is designed and built by MPL and FAU. It is based on fiber-based delay-line interferometers and superconducting nanowire single-photon detectors [Fig. \ref{fig:E2E_Phase}(c)] \cite{hacker_phase-locking_2023,gunthner_eagle-1_2024}. To facilitate the key exchange, the end user communicates over public channel via the OGS and satellite OST with the QPL. The end user can be located remotely from the OGS, where the maximum distance is limited by fiber loss.

\section{Conclusion}
With the 2026 satellite launch, EAGLE-1 will be Europe's first satellite-based end-to-end QKD system. It represents an important technological milestone with its novel all-optical C-band approach in combination with phase-encoded double pulses as qubits. The mission also underlines Europe's ambition to take a leading role in the field of quantum technologies and its commercial applications and paves the way towards a European QKD satellite constellation which constitutes an integral part of Iris\textsuperscript{2} and the European Quantum Communication Infrastructure (EuroQCI) initiative\cite{noauthor_european_2024}.  

\acknowledgments   
We thank the German Space Agency (DLR), ESA and the European Commission for their continued support and  assistance. We acknowledge the constructive and collaborative partnership with SES and all consortium members. Finally, we are grateful for the motivation, commitment and engagement of all TESAT colleagues that are contributing to the success of EAGLE-1, especially Philipp Hagel, Felix B\"{u}rger and Hannes Hoos.

\bibliographystyle{spiebib}

\end{document}